\documentclass{aa}
\usepackage{graphicx}
\usepackage{natbib}

\begin{document}

 \title{Relativistic spine jets from Schwarzschild black holes:}
  \subtitle{Application to AGN radioloud sources}


  \titlerunning{Relativistic spine jets}

 \author{Z. Meliani,
          \inst{1,2}
  \and  C. Sauty
          \inst{2}
   \and  K. Tsinganos
          \inst{3}
   \and  E. Trussoni
          \inst{4}
  \and  V. Cayatte
          \inst{2}
         }

   \offprints{
{\tt christophe.sauty@obspm.fr}}
  \institute
{       Centrum voor Plasma Astrofysica, Celestijnenlaan 200B bus 2400, 3001 Leuven, Belgium
   \and
       Observatoire de Paris, LUTh., F-92190 Meudon, France
   \and
       IASA and Section of Astrophysics, Astronomy \& Mechanics Department
       of Physics, University of Athens, Panepistimiopolis GR-157 84, Zografos, Greece
   \and INAF - Osservatorio Astronomico
      di Torino, Via Osservatorio 20, I-10025 Pino Torinese (TO), Italy}
  \date{Received ... / accepted ...}


 \abstract
{The two types of Fanaroff-Riley radio loud galaxies, FRI and FRII, exhibit strong jets but with
different properties. These differences may be associated to the central engine and/or the external medium.}
{The AGN classification FRI and FRII can be linked to the rate of electromagnetic Poynting flux
extraction from the inner corona of the central engine by the jet.   The collimation results from the distribution of the total electromagnetic energy across the jet, as compared to the corresponding distribution of the thermal and gravitational energies.}
{We use exact solutions of the fully relativistic magnetohydrodynamical (GRMHD) equations obtained by a nonlinear
separation of the variables to study outflows from a Schwarzschild black hole corona.}
{A strong correlation is found between the jet features and the energetic distribution of the plasma of
the inner corona which may be related to the efficiency of the magnetic rotator.}
{It is shown that observations of FRI and FRII jets may be partially constrained by our model for spine jets.
The deceleration observed in FRI jets may be associated with a low magnetic efficiency of the central
magnetic rotator and an important thermal confinement by the hot surrounding medium.
Conversely, the strongly collimated and accelerated FRII outflows may be self collimated by their own magnetic field because of the high efficiency of the central magnetic rotator.
}


\maketitle
%

\section{INTRODUCTION}

{ According to the standard Active Galactic
Nuclei (AGN) paradigm, their radio luminosity is related to the 
presence of powerful relativistic jets (radio loud AGN),
or, to mildly sub-relativistic outflows (radio quiet AGN). 
And, by assuming a supermassive BH surrounded by an accretion disk/torus, the different
AGN phenomenologies observed in both classes are related to the orientation
of the axis of the BH/disk system with respect to the line of sight, and the thickness of the torus 
which is responsible for the obscuration effects \citep{UrryPadovani95}. 
Typical examples for radio quiet AGN are the various types of Seyfert I - II galaxies, where uncollimated (or
loosely collimated) winds are outflowing from the BH/disk system at a speed of a few thousands km s$^{-1}$.
However, besides their inclination to the line of sight the classification cannot be complete without 
invoking another key parameter to explain the outflow differences between the various AGN. These
differences may be related to galaxy environment effects and/or intrinsic properties of the
AGN, as shown in Fig. 1  (see e.g. Kaiser \& Alexander 1997; Celotti 2003;
 Kaiser \& Best 2007).

For radio loud objects a fundamental role is played by Doppler boosting,
strongly affecting the luminosity and spectral properties of these
AGN. In fact,  these radio sources are associated with powerful relativistic jets which reach 
at the parsec scale, high Lorentz factors $\gamma\sim 5-30$ \citep{UrryPadovani95, Kellermannetal04, Pineretal03}.
  Such jets are strongly collimated with opening angles of the order of 
$\sim 3^{\circ}$ \citep{Pushkarevetal09}. In nearby AGN recollimation is 
inferred from the inner radio jet structure (Horiuchi et al. 2006 for Cen A; 
Kovalev et al. 2007 for  M87).
We recall that the main classes of radio loud AGN are Radio Quasars,
Flat Spectrum and Broad Line Radio Galaxies, BL Lacs, Fanaroff Riley I
(FRI) and  Fanaroff Riley II (FRII) objects. According to the unified model,
FRI objects are misaligned BL Lacs, while the parent population of  FRII are
Radio Quasars, Broad Line Radio Galaxies and the brightest BL Lacs, as sketched in Fig. \ref{figAGNclassif}.
Regarding in particular the FR I and FR II dichotomy, we briefly outline in the following their 
main properties \citep{Fanaroff&Riley74}:

\noindent
- In FR II sources the extended radio morphology shows a clear, generally one sided collimated
 (within a few degrees) thin jet, terminating into a hot spot and surrounded by diffuse blobs. 
 Conversely, in FR I sources the collimated symmetric jets smoothly merge into the extended emitting regions.

\noindent
 - FR II jets look highly relativistic and narrow  along their
 whole length (tens of kpc). FRI jets are conversely relativistic only on pc scales \citep{Bridle92},
 becoming subrelativistic  and diffuse on kpc scales \citep{Giovanninietal05}.
However, in some FR I sources the structure of the jet in the kiloparsec scale appears
more complicated, with an inner spine that remains
relativistic and an outer shell that decelerates and becomes sub-relativistic \citep{Canvinetal05}.

\noindent
- FR II sources are more powerful than the FR I ones, with threshold power $\sim 10^{25}$ W
 Hz$^{-1}$ sr$^{-1}$ increasing with the radio galaxy luminosity
(Ledlow \& Owen 1996).

\noindent
- FR II are usually found in poor gas environments, with jets probably
 collimated by their helical magnetic fields
\citep{Hardcastle&Worrall00, Asadaetal02, Zavala&Taylor05}
 and slightly interacting with the
 external gas. Rich environments harbor mostly FR I sources and their jets, thermally
 confined (at least partially) and appearing to strongly interact with the
 intracluster medium \citep{KaiserAlexander97, Laingetal99, Gabuzda03}.
 The measured transverse magnetic field suggests the presence of internal
 shocks where the tangled magnetic field is compressed \citep{Gabuzdaetal94, Gomezetal08}.
   Those shocks could be the result of the thermal collimation of the jet.
In the following, we briefly discuss numerical versus analytical modeling of multicomponent jets. 

 There are two main theories to interpret the above observational characteristics.
The first explains the morphological
differences as mainly due to the different physical properties of the environment
in which the relativistic jet propagates
\citep{DeYoung93, Bicknell95, Laingetal99, Gopal-KrishnaWiita00,Melianietal08}.
The second explains the dichotomy by involving a difference in the
nature of the central engine,  the spin of the central black
hole, the accretion rate and the jet composition \citep{Baumetal95, Reynoldsetal96, Meier99, MelianiKeppens09}.
Finally, there may be a combination of external and engine
factors to explain the FR I/FR II dichotomy, as we have suggested in Meliani et al. (2006a)
and studied in Wold et al. (2007).}


Similarly to jets from Young Stellar Objects \citep{Ferreiraetal06}, AGN
jets probably have at least two components \citep{Soletal89, TsinganosBogovalov02} one
originating from the surrounding Keplerian disk
(Baum et al. 1995; Meier 2002; Begelman \& Celotti 2004)
and the other from
the inner corona surrounding the central black hole. This corona can
be created  by the "CEntrifugal pressure supported Boundary
Layer" model (CENBOL) \citep{DasChakrabarti02}.
{ The corona can also be created  by the mechanism
presented in \cite{Kazanas&Elison86}.  A third alternative to produce such a
corona with pair plasma is the Blandford \& Znajek (1977) model where the jet
is powered by the spinning black hole. However observations indicate that the
jet should have both hadronic and leptonic components as explained in two-component 
models (Henry \& Pelletier 1991;  Fabian \& Rees 1995).} 

{ The MHD equations can be solved through numerical simulations, which describe
the evolution of the
jet configuration. The availability of more and more powerful computing
facilities and sophisticated numerical codes allows a quite complete
description and understanding of the jet acceleration/collimation
\citep{Komissarovetal07, Porth&Fendt10} and the accretion/ejection process
\citep{Koideetal98, Koideetal99, McKinney06, McKinneyBlandford09,
 Graciaetal06, Graciaetal09}. Recent numerical simulations have progressed to
general relativistic magnetohydrodynamic (GRMHD) jet launching, as in
\cite{McKinney06} and \cite{Hardeeetal07},  suggesting also the formation of
jets with two components.
However computational limits do not allow yet to follow simulations for very long times
and reach exact stationary configurations. It also fails at analyzing
structures with very different scale lengths.

\hspace{-0.5cm}\begin{figure}
\includegraphics[height=8cm]{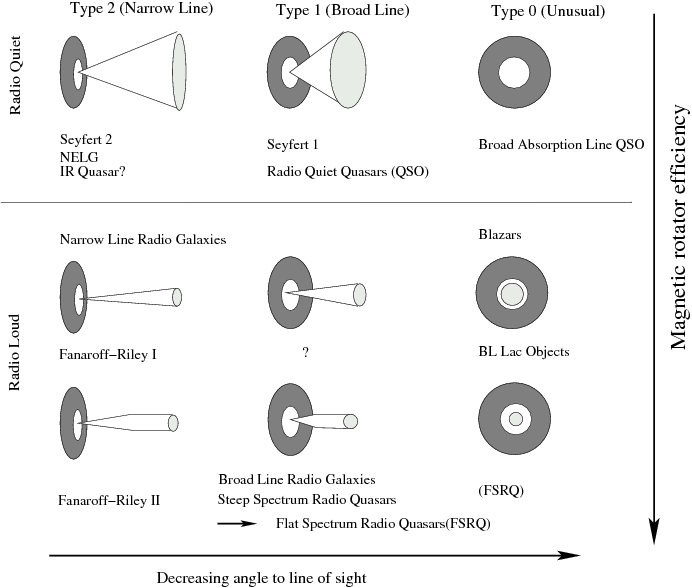}
\caption{Standard classification of AGN sources following Urry \&
Padovani (1995). The horizontal axis represents 
the inclination of the source axis with the line of
sight. The vertical axis we suggest that it may be linked to the
efficiency of the underlying magnetic rotator to collimate the
flow.}
\label{figAGNclassif}
\end{figure}

Nevertheless, tremendous progress on understanding the physics of relativistic
jet acceleration/deceleration -- and therefore the FRI/FRII dichotomy -- has
been done thanks to numerical simulations of jet propagation in the asymptotic
regions.
Some investigated the relativistic hydrodynamic jet propagation through the
interstellar medium
\citep{DuncanHughes94, Martietal97, KomissarovFalle98, Aloyetal99, Rossietal08}.
They show that the different dynamics of FR I and FR II jets may be a
consequence of the power of the jet.
Many groups had also investigated the  two-dimensional relativistic magnetized
jet propagation in an external medium
\citep{vanPutten96, Komissarov99, Leismannetal05, Keppensetal08}
and 3D
(e.g. Mizuno et al. 2007, Mignone et al. 2009). They showed that the
interaction between the jet and the external medium depend on the
magnetization of the jet and the density ratio between the jet and the
external medium. The role of the environment
may be crucial in HYbrid MOrphology Radio Sources~
(HYMORS), as shown by
Gopal-Krishna \& Wiita (2000, 2002). These radio sources appear to have
a FR II type on one side and a FR I type diffuse radio
lobe on the other side of the active nucleus.
This last model for HYMORS has been recently confirmed by numerical
simulations of two component jets \citep{Melianietal08}. Another alternative to
the FRI/FRII dichotomy could come from the nature of the instabilities that
develop in the jet, or, at the jet interface with the external medium, or, with
another surrounding outflow component \citep{Keppensetal08}. This is
consistent with the fact that hydrodynamical jets with high Lorentz factors
are more stable \citep{Martietal97}. Poloidal magnetic fields also help to
stabilize the jet \citep{Keppensetal08}. On the last alternative note that simulations by
\cite{MelianiKeppens09} confirm that the deceleration in FRI jets could be
attributed to a strong Rayleigh Taylor instability between the spine jet and
the surrounding component, supporting a two component jet structure.}

{ In parallel to the development of time dependent simulations,
steady jet solutions in GRMHD were first obtained numerically by solving the
transfield equation in the force-free limit (Camenzind 1986). This study has
been further
developed in GRMHD by using first a Schwarzschild metric and then extending it
to a Kerr metric (Fendt 1997). However this method cannot incorporate
consistently the mass loading of the jet.

Exact models for disk winds can be constructed via a nonlinear separation of
the governing MHD equations. This technique of radial self-similarity has been
largely explored in the context of stellar and relativistic jets by various
authors \citep{BardeenBerger78, BlandfordPayne82, Lietal92, Contopoulos94,
 Ferreira97,  VlahakisKonigl04}.

Meridional self similarity is another way to variable separation which is used to produce
models of pressure driven winds \citep{Melianietal06}. It is also
complementary to magnetically driven disk winds. Such models may describe the
inner spine jet from the central object necessary to sustain the outer disk
wind but where the radial self-similar models fail by construction. In this
paper, using such solutions, we show that the collimation criterion developed
in the frame of this model can help understanding how the FRI/FRII dichotomy
may influence the morphology of the inner  spine jets. As a back reaction, the
spine jet dynamics influences the outer jet as it was demonstrated in
numerical simulations \citep{Melianietal06b, Fendt09}, even if the central jet
is energetically very weak \citep{Matsakosetal09}.}

In Sect. \ref{BasicEq} we recall briefly the main assumptions of the
model. In Sect. \ref{Jetclassif} we summarize the details of the
standard AGN jet classification and how it helps to constrain the
parameters. In Sect. \ref{FRI} we present an interesting solution
for FRI type spine jet and in Sect. \ref{FRII} another one for jets associated with FRII
objects. In Sect. 6 we discuss and summarize the main implications of our model.

\section{Model assumptions}\label{BasicEq}
\subsection{The self-similar model}
{
We use the ideal GRMHD equations in the background spacetime of a Schwarzschild
black hole, and neglect the effects of self gravity of matter outside the black
hole. The spacetime curvature at a distance $r$ from the black hole is given
by the  lapse function,
\begin{equation}
h =\sqrt{1-\frac{r_{\rm G}}{r}}\,,
\end{equation}
where $r_{\rm G}$ is the Schwarzschild  radius.

Following Meliani et al. (2006a), all physical quantities are normalized
at the Alfv\'en radius $r_\star$ along the polar axis where the meridional
angle is zero ($\theta=0$). We define a dimensionless spherical radius
$R=r/r_\star$, cylindrical  radius
$G$, and magnetic flux function $\alpha$,
\begin{equation}
\alpha=\frac{R^2}{G^2(R)}\sin^2 \theta\,, \qquad G=\frac{r\,\sin(\theta)}{r_\star\,\sin(\theta_\star)}
\,.
\end{equation}

To describe the GRMHD outflow of the coronal plasma, we use the relativistic meridionally
self-similar solutions presented in Meliani et al. (2006a).
The specific enthalpy, and density in the lab frame, together with the pressure, velocity and
magnetic field are given in terms of functions of the radial distance $R$,
\begin{eqnarray}
\label{hgw}
h\gamma w &=& h_{\star} \gamma_\star w_\star
\left(1-\frac{\mu \lambda^2}{\nu^2}
\frac{N_B}{D} \alpha \right)\,,\\
\label{hgn}
h\gamma n &=& h_{\star} \gamma_\star  n_\star
\frac{h_{\star}^2}{M^2}\left(1+\delta\alpha -\frac{\mu
\lambda^2}{\nu^2} \frac{N_B}{D} \alpha
\right) \,,\\
\label{pressure}
P&=&P_0 + \frac{1}{2}{\gamma_{\star}}^2n_{\star}\frac{w_\star}{c^2}
V_ {\star}^2\Pi(R)(1+\kappa\alpha)\,,\\
\label{Vr}
V_r&=&\frac{V_\star M^2}{h_{\star}^2G^2}
\frac{1}{\sqrt{1+\delta\alpha}}
\left(\cos{\theta}
+\frac{\mu \lambda^2}{\nu^2}\frac{N_B}{D} \alpha
\right) \,,\\
\label{VQ}
V_{\theta}&=&-\frac{V_\star M^2}{h_{\star}^2G^2}
\frac{h_{}F}{2}\frac{1}{\sqrt{1+\delta\alpha}} \sin{\theta}
\,,\\
\label{Vphi}
V_{\varphi}&=&-\frac{h_{} }{{h_{\star}}}
\frac{\lambda V_\star}{G^2}
\frac{N_V}{D}
\frac{R\sin{\theta}}{\sqrt{1+\delta\alpha}}\,,\\
\label{Br}
B_r&=&\frac{B_\star}{G^2}\cos{\theta}\,,\\
\label{BQ}
B_{\theta}&=&- \frac{B_\star}{G^2}\frac{h_{}F}{2}\sin{\theta}\,,\\
\label{Bphi}
B_{\varphi}&=&-\frac{\lambda B_\star}{G^2}
\frac{h_{\star}}{h_{}}
\frac{N_B}{D} R\sin{\theta}
\,.\end{eqnarray}
where
\begin{eqnarray}
N_B=\frac{h_{}^2 }{h_{\star}^2} - G^2\,,\;
N_V=\frac{M^2}{h_{\star}^2}-G^2\,,\;
D=\frac{h_{}^2 }{h_{\star}^2}-\frac{M^2}{h_{\star}^2}
\,.\end{eqnarray}
The free parameters  $\delta$ and $\kappa$ describe the deviation
from spherical symmetry of the ratio of number density/enthalpy
and pressure, respectively, while $\lambda$ is a constant controlling the angular momentum
extracted by the jet.
The constants $\nu$ and $\mu$ measure the escape speed in units of the light speed
and the escape speed in units of Alfv\'en speed, at the Alfv\'en point along the polar axis, respectively,
\begin{equation}\label{munu}
\mu=\frac{V_{{\rm esc,}\star}^2}{c^2}\,, \qquad \nu=\frac{V_{{\rm esc,}\star}}{V_\star}
\,.
\end{equation}

Thus, all physical quantities are determined in terms of constant parameters, $\delta$, $\kappa$,
$\lambda$, $\mu$, $\nu$ and the
three unknown functions, $\Pi(R)$, $F(R)$ and $M^2(R)$. Note that $\Pi (R)$ is the dimensionless
pressure function, defined modulus a constant $P_0$, $F$ is the expansion factor,
and $M$  is the poloidal Alfv\'enic number,
\begin{equation}\label{FM}
F=2-\frac{{\rm d}\ln{G^2}}{{\rm d}\ln{R}}\,,\qquad M=\frac{4\pi h_{}^2 n w
 \gamma^2 V_{\rm p}^2} {B_{\rm p}^2 c^2}
\,.
\end{equation}
These three unknown functions $\Pi(R)$, $F(R)$ and $M^2(R)$ are determined by
three nonlinear equations. We start integrating these equations from the
Alfv\'en critical surface, taking into account there the corresponding
regularity condition and then integrate downwind and upwind, crossing the
other critical points, for details see Meliani et al. (2006a).

{ The light cylinder is  defined by the function $x=\Omega\, L/ E$ becoming
 unity, where $\Omega$ is the angular
speed, $L$ the total angular momentum per unit mass and $E$ the generalized Bernoulli
integral. It is  a measure of the energy flux of the magnetic rotator in units
of the total energy flux. The s-s description is possible only if the jet is rotating at
subrelativistic speeds. In such conditions $x$ must remain small and therefore
the light cylinder effect have to be negligible for our solution to be valid.}


{ In this setup of the relativistic MHD problem an extra
parameters exists $\epsilon$ which is constant everywhere 
\citep{Melianietal06}:
\begin{eqnarray}\label{varepsilon_full}
\epsilon =
\frac{M^{4}}{h_{\star}^4 R^2 G^2}
\left(\frac{F^2}{4} - \frac{1}{h^2} - \kappa \frac{R^2}{h^2G^2}\right)
-\frac{\left(\delta-\kappa\right)\nu^2}{h^2 R}
\nonumber \\
+\frac{\lambda^2}{G^2 h_{\star}^2} \left(\frac{N_V}{D}\right)^2
+\frac{2 \lambda^2}{h^2}\frac{N_B}{D}
\,.
\end{eqnarray}
This is the relativistic generalization for a Schwarzschild black hole
of the classical constant found in Sauty et al. (2004)
that  measures the magnetic energy excess or deficit on a nonpolar streamline, compared to
the polar one.} To first order, $\epsilon$  determines the fraction of
Poynting flux carried by the jet in the asymptotic region, such that $1-\epsilon$
measures the fraction of the Poynting flux  which is used in the jet acceleration.
Thus, if $\epsilon>0$ we have an Efficient Magnetic Rotator (EMR) where
magnetic collimation may dominate, while if $\epsilon<0$ we have an
Inefficient Magnetic Rotator (IMR) where collimation cannot be of magnetic
origin  but should be thermal, if the flow is collimated.
}

\section{New interpretation of the Fanaroff-Riley classification}\label{Jetclassif}

We propose here to examine the vertical classification in
Fig. \ref{figAGNclassif} by means of our parameter $\epsilon$,
or less ambitiously, to interpret how the various observations we have
on FRI and FRII jets may influence the formation of the inner spine
jet component. As in the non relativistic
case \citep{STT99}, this parameter $\epsilon$ allows to classify jets according to the
efficiency of the central magnetic rotator.  Jets emerging from efficient central magnetic
rotators, $\epsilon >0$, collimate cylindrically without
oscillations in the asymptotic region. In this type of jets, the
velocity increases monotonically to reach its asymptotical maximum value. Conversely, jets associated
with inefficient central magnetic rotators, $\epsilon <0$, are collimated mainly by the
pressure of the external medium. Thus, this type of jets strongly
interacts with the ambient medium and this induces oscillations in
their shape at the asymptotic region. The speed of these jets also does not increase
monotonically but it oscillates too. The plasma in those jets is
accelerated until an intermediate region where the speed reaches its maximum value.
Then, the acceleration of the jet stops in this region. Further away in the recollimation region
the outflow slows down. Note that such jet oscillations could lead finally to a more turbulent outflow
consistently with the numerical simulations of two component jets we mentioned
in introduction. After the recollimating region we cannot exclude the presence of shocks or instabilities. Thus it is difficult to know if the oscillations that appeared in the solutions would be observable or not.

In summary, we propose that a difference between these two types of
spine jets  associated with FR I and FR II radio galaxies may result
from the competition between the magnetic  and thermal confining mechanisms.
Magnetic pinching and pressure gradient tend
to compensate the transverse expansion of the jet because of the
centrifugal force and the charge separation which also induces an
outwards electric force. { These two expanding forces are
characterized by the free parameter $\lambda^2$, that
measures the quantity of angular momentum
carried along the streamlines.} Magnetic collimation is controlled by
the parameter $\epsilon/\lambda^2$ (see Sauty et al. 1999).  On the other hand,
thermal collimation is controlled by the parameter $\kappa$ that
defines the transverse variation of the pressure.

Hence, an appropriate manner to classify the different jet solutions
according to the nature of their collimation is based on the two
free parameters $\kappa/2\lambda^2$ and $\epsilon/2\lambda^2$
\citep{STT99, Sautyetal2002, Sautyetal04}. The higher is the value
of those parameters, the stronger is the collimation and the lower
is their terminal speed. This can be seen in Tab. \ref{Tab1} where
we plot for various solutions  the asymptotic jet speed
in units of the escape speed from the base of the corona: as the
efficiency of collimation increases, the efficiency of
acceleration decreases.  In other words, tightly collimated jets
(larger values of $\epsilon/2\lambda^2$, or, $\kappa/2\lambda^2$)
have lower terminal speeds.
{Thus the strong interaction with the external medium would naturally result
into a decelerated flow even if it remains stable. Parallely, the jet being
denser  would radiate more on large scale before the terminal shock with the ambient medium}.

\begin{table*}
\begin{center}
\begin{tabular}{ccccccc}
  \hline
  \hline
 $\frac{\epsilon}{2\lambda^2}$/$\frac{\kappa}{2\lambda^2}$  & $-0.5$& $-0.05$ &
$-0.5\,10^{-2}$& $0.5\,10^{-2}$& $0.05$  & 0.5\\
\hline
 $0.5\,10^{-1}$ &3.4& 1.76& 1.59& 1.3 & 1.07  & 0.847 \\
 $0.5\,10^{-2}$ &30 &13  & 7.0 & 5.77& 1.069 & 1.66\\
$0.5\,10^{-4}$  &90 &75  & 65  & 60  & 10    & 1.02 \\
\hline
\end{tabular}
\label{Tab1}
\caption{The asymptotic jet speed in units of the escape speed from
the base of the corona,  for various values
of the parameters $\kappa/2\lambda^2$ and $\epsilon/ 2\lambda^2$}
\end{center}
\end{table*}

\begin{figure}[h]
{\rotatebox{0}{\resizebox{9.3cm}{5.5cm}
{\includegraphics{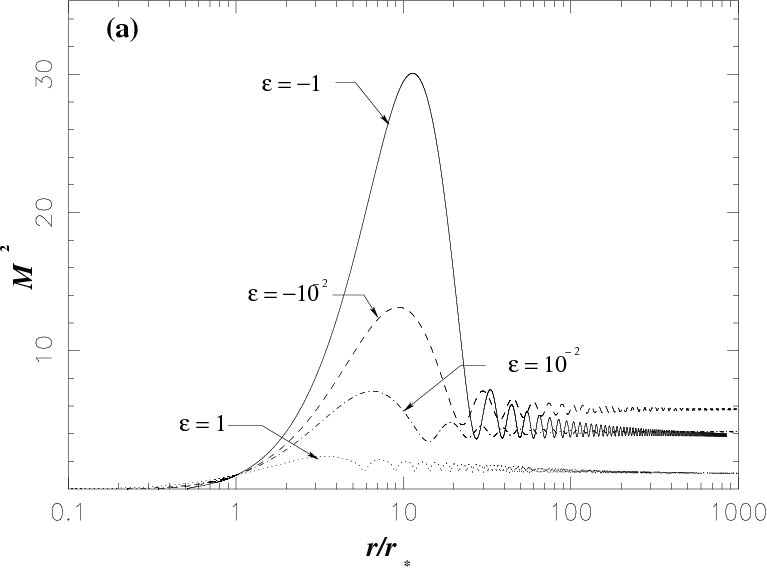}}}}
\rotatebox{0}{\resizebox{9.3cm}{5.5cm}
{\includegraphics{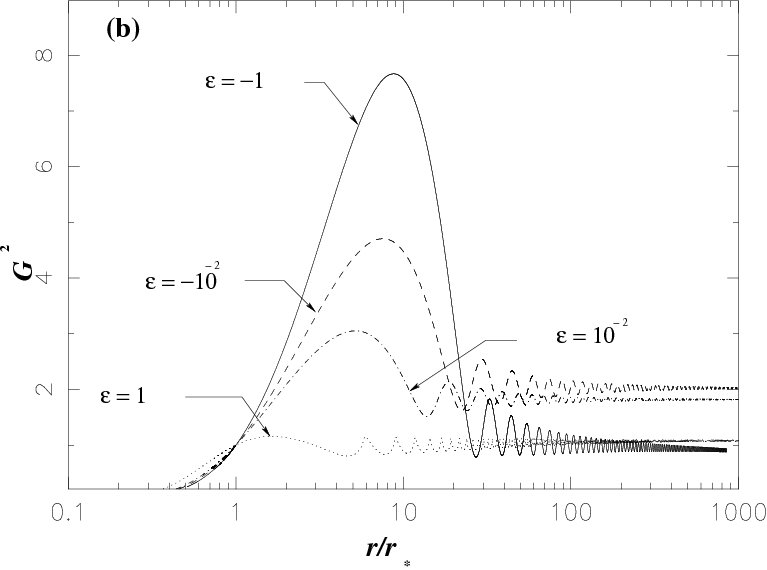}}}
\caption{ In (a) we plot the Alfv\'en number $M^2$ and, in (b), the cylindrical jet
cross section  $G^2$ as functions of the distance for  $\kappa/2\lambda^2 = 0.05$ and
four values of $\epsilon/2\lambda^2$ $(-0.5, -0.5\times 10^{-2}, 0.5\times 10^{-2},0.5)$,
corresponding to Tab. \ref{Tab1}.}
\label{epsilonkappaAGN}
\end{figure}

Tab. \ref{Tab1} also shows that jets from inefficient
magnetic rotators are more powerful in transforming thermal energy
into kinetic energy, than those from efficient magnetic rotators.
There are two reasons for this.

{\it First}, { in a EMR flow, the centrifugal force
at the base of the jet is important. Then, the last stable orbit of the
plasma gets closer to the central black hole.  Therefore, the corona extends
closer to the black hole horizon. 
} 
Thus, if the
available total amount of thermal energy is the same in the corona, the
plasma gravitational potential increases. Consequently, in EMR as
$\epsilon/ (2\lambda^2)$ increases, more thermal energy is tapped
in order to allow for the plasma to escape and less is left for accelerating
it.

{\it Second} (see Fig. \ref{epsilonkappaAGN}), an increase of the
magnetic rotator efficiency limits the initial expansion of the
outflow. The pinching magnetic force gets stronger after the
Alfv\'en surface and the conversion of thermal energy into kinetic
energy stops when the jet reaches its asymptotic cylindrical shape.
This decrease of the jet asymptotic speed with the increase of the
magnetic rotator efficiency may seem contradictory with the usual picture.
However, in axial outflows the contribution of the Poynting flux to
the total acceleration remains weak. This is of course different
from relativistic disk wind models where the acceleration is
dominated by the conversion of Poynting flux to kinetic energy flux
\citep{Lietal92, Contopoulos94,
VlahakisKonigl04}. Those jets  are characterized by a strong
inclination of the magnetic field lines and the rotation at the base
of the outflow is almost Keplerian.
Note also that, as expected, the efficiency of the acceleration also
increases with the degree of expansion, i.e., as $\kappa/2\lambda^2$ decreases.

\subsection{Estimation of the free parameters}

In the following sections, we mainly explore two examples of typical
solutions for relativistic jets. One is associated with an IMR such
that the contribution of the external pressure is comparable to the magnetic
one, while the
second is associated to an EMR such that the jet is self collimated
magnetically. In order to find these solutions we first make an
estimation of the free parameters of the model
using our knowledge of the properties of FRI
and FRII jets, in particular in the launching region on one hand and
in the far asymptotic region on the other hand.

At the base of the outflow, we assume that the corona starts at
the radial distance of the last stable orbit which is around the
radius of the Schwarzschild black hole, { as discussed below}.
We consider that the last open streamlines, which emerge from the
corona, should be in sub-Keplerian rotation if they are anchored in the
thick disk surrounding the central black hole. The last open
streamline in the coronal jet is the one that crosses the
equatorial plane at the edge of the magnetic dead zone, which we assume to have a
dipolar configuration.

{ We use also some observational constraints in the asymptotic
region of the jet. The opening angle of jet has to be a few degrees.
Six degrees is the value inferred
for the well measured jet of M87 (Biretta et al. 2002;
Kovalev et al. 2007). We guess that the spine jet
opening angle is even smaller and we took a value of $\simeq 1^{\circ}$ at 1 pc.
Then, from the expression of the dimensionless magnetic
flux  $\alpha$, by knowing the last streamline,
we can deduce the asymptotic value of $G_{\infty}$ :
\begin{eqnarray}\label{Jet_1_G_infty_DET}
G_{\infty} = \frac{r_{\infty}}{r_{\star} \sqrt{\alpha_{\rm ext}}}
\sin{\theta}\,,
\end{eqnarray}
where $\alpha_{\rm ext}$ is the last open streamline in the jet. As a reasonable
estimate, we have taken $\alpha_{\rm ext} = 4$. In fact, for smaller
values of $\alpha_{\rm ext}$, the value of $G_{\infty}$ becomes too large
and therefore unrealistic for our model.
We further assume that in the asymptotic region the Alfv\'en number
is of the order of $M_{\infty}\simeq  5$:} { this is the order
of magnitude found in the literature for relativistic magnetohydrodynamical
jet propagation (Leismann et al.  2005, Keppens et al. 2008).}

{ We know that the asymptotic Lorentz factors should be
   between $\sim$ 3 and 10 \cite[see][]{Pineretal03}. From the expression of the asymptotic
velocity  along the polar axis (Eq. \ref{Vr}) we can
deduce from Eq. 13 the value of the free parameter $\nu^2$, }
\begin{eqnarray}\label{Jet_1_nu_DET}
\nu^2 = \frac{\mu c^2}{V_{\infty}^2}
\left(\frac{M^2_{\infty}}{h_{0\star}^2 G^2_{\infty}} \right)^2.
\end{eqnarray}

We used on purpose a rather lower limit for the Lorentz factor in order to avoid
large effective temperatures in our model. In fact, we can use the same solutions and scale them up to
obtain higher Lorentz factors but then the effective temperature would attain extremely large values
above the mass temperature. However the corresponding high pressure could have a large contribution
from a turbulent magnetic or ram pressure component in the jet (see Aib�o et al. 2007, for the solar wind).
In such a case the kinetic temperature would be lower.
Nevertheless, as far as the collimation is concerned this does not affect qualitatively our discussion
on the dichotomy between FRI and FRII and we kept this relatively low Lorentz factor.

We used also the observed mass loss rate in the outflow to constrain
the free parameter $\delta$. However, the spine jet probably carries only a small
fraction of the observed energy flux in AGN jets, $L_{\rm jet,Kin} \sim 10^{43} \,{\rm ergs/s}$ \citep{Allenetal06}
The energy flux  of the coronal wind
remains weak compared to the total mass carried by the disk-wind
which is supposedly denser (Vlahakis \& Konigl 2004), a situation similar to stellar jets
associated with Young Stellar Objects \citep{Melianietal06b}.
Then $\delta$ is deduced from the assumed value for the mass loss rate,
\begin{eqnarray}
\dot{M}&=&2\,\int_{\rm section} m_{\rm part}\,h\,\gamma\,n\,V_{p}\, {\rm d}S\,,\nonumber\\
&=&\,\frac{4\,\pi\, m_{\rm part }r_{\rm G}^2}{\mu^{3/2}\,\nu}\sqrt{\gamma_{\star}^2\, n_{\star}\,{w_{\star}/c^2}}\, \int_{0}^{\alpha_{\rm ext}}\sqrt{\frac{M^{2} n}{w/c^2}}{\rm d}\alpha\,.
\end{eqnarray}
We get an equation for $\delta$,

\begin{eqnarray}\label{Jet_1_delta_DET}
&&(1 + \alpha_{\rm ext} \delta)^4\times C^2  -(2  C^2+1 ) \times(1 + \alpha_{\rm ext} \delta)^3  +\nonumber\\
&& (1 + \alpha_{\rm ext} \delta)^2\times C \times (2 + C) \nonumber\\
&&- 2 \times C\times (1 + \alpha_{\rm ext} \delta) + 1 = 0\,,
\end{eqnarray}
where $C$ is another constant given by,
\begin{eqnarray}\label{Jet_1_AA_DET}
C = \frac{3{\mu^{3/2}\nu  (1 - \mu)}\dot{M}}{8 \pi r_{G}^2 m_{\rm part}{\sqrt{1-
r_G/r_{\infty}}} \gamma_{\infty}
n_{\infty} M^2_{\infty}\alpha_{0}}
\,.
\end{eqnarray}
The variable  $n_{\infty}$ is the asymptotic density and $m_{\rm par}$
the average mass of the particles. We consider a proton-electron fluid, $m_{\rm par} = m_{\rm proton}$.

{ For the mass loss rate, we choose
$\dot{M} =  n_{0} 10^{-6}\dot{M}_{\rm Edd}$, where $\dot{M}_{\rm Edd}$
is the Eddington mass loss rate. It
corresponds to value  found for the
relativistic Parker wind \citep{Melianietal04}. } The asymptotic density is taken equal to
$n_{\infty} = 10^{-8}\times n_{0}$ cm$^{-3}$ \citep{Melianietal04}, with $n_{0}$
a dimensionless free parameter.

We suppose that the rotation is sub-Keplerian on the last open
streamline of the jet at the equator. The parameter $\eta$
measures the deviation of the rotation function
$\Omega$ from its Keplerian value,
\begin{eqnarray}\label{Jet_1_Omega_subkepl}
\Omega = \left(1 - \eta\right) \sqrt{\frac{r_{G}^2}{2 r_{0}^3 \mu}}
c\,,
\end{eqnarray}
{ From the definition} of $\Omega$
(Eq. \ref{Jet_1_Omega_subkepl}) we can deduce the value of the free
parameter $\lambda$ which is the constant controlling the angular momentum extracted by
the jet,
\begin{eqnarray}
\lambda& = &\left(1 - \eta\right) \sqrt{\frac{r_{G}^2}{2 r_{c}^3
\mu}} \frac{r_{\star} \sqrt{1 + \delta \alpha_{0}}}{h_{0 \star}
\frac{v_{\star}}{c}}\nonumber\\
& = &\left(1 - \eta\right) \sqrt{\frac{r_{G}^2}{2 r_{c}^3 \mu}}
\frac{r_{\star} \sqrt{1 + \delta \alpha_{0}}}{h_{0 \star} }
\sqrt{\frac{\nu}{\mu}}.
\end{eqnarray}

As we mention earlier we assume that the corona forms above the last stable orbit at
$r_{0} = 3 r_{G}$. We choose a typical magnetic lever arm (i.e. Alfv\'en radius) of 10 times the
Schwarzschild radius, $r_\star=10r_G$. It gives,
\begin{eqnarray}\label{Jet_1_Position_couronne_DET}
R_{0} = \frac{r_{0}}{r_{\star}} = 0.3 \,.
\end{eqnarray}
The parameter $\kappa$, which is the relative variation of the pressure with
latitude, is calculated from an approximate expression of the coronal
base  where $M_0 \rightarrow 0$ and $dM^2_0/dR$ is finite in Eq. A.2 - A.5 (Meliani et al. 2006a) using,
\begin{eqnarray}
\kappa = \delta - R_{0}  \frac{2 \lambda^2}{\nu^2} \left(1 -\mu\right)
\,.
\end{eqnarray}

To summarize, our free parameter estimates are:
\begin{enumerate}
\item The Alfv\'en radius
\begin{eqnarray}
r_{\star} = 10\times r_{G} \rightarrow \mu =
0.1 \,.
\end{eqnarray}
\item The jet opening angle at
$r_{\infty} \simeq 1$ pc is $\simeq 1^{\circ}$ which gives
the asymptotic value of $G_{\infty}$ for the last streamline
$\alpha_{\rm ext} = 4$.
\item { For the asymptotic Alfv\'en number we chose $M_{\infty}\sim 5$.}
\item The asymptotic Lorentz factor is taken to be $\gamma_{\infty} \sim
2$.
\item
The asymptotic density is taken
equal to $n_{\infty} = n_0 \times 10^{-8}$ cm$^{-3}$.
\item The new parameter $\eta$ which measures how sub-Keplerian is the velocity (the values
differ from solution to solution).
\item The corona is supposed to be formed above the last stable orbit,
$r_{c} = 3 r_{G}$ which gives $\kappa$.

\end{enumerate}

{ The parameter $\eta$ is not independent from the rest of the
model. However it measures precisely the rotation of the footpoints and controls the
efficiency of the magnetic rotator to collimate the jet. Thus we
have an indirect way to determinate the properties of the disk from the
asymptotic characteristics of the jet. If $\eta \rightarrow 0$ open
streamlines anchored in the accretion disk rotate with an almost Keplerian
profile. On the other hand, if $\eta \rightarrow 1$ open streamlines anchored
in the accretion disk rotate very slowly  and the Poynting flux injected in
the jet is low.}

Before going further into our modeling of FRI and FRII jets, we
need to  introduce the notion of effective temperature.
\begin{figure}[h]
{\rotatebox{0}{\resizebox{9.5cm}{9.5cm}
{\includegraphics{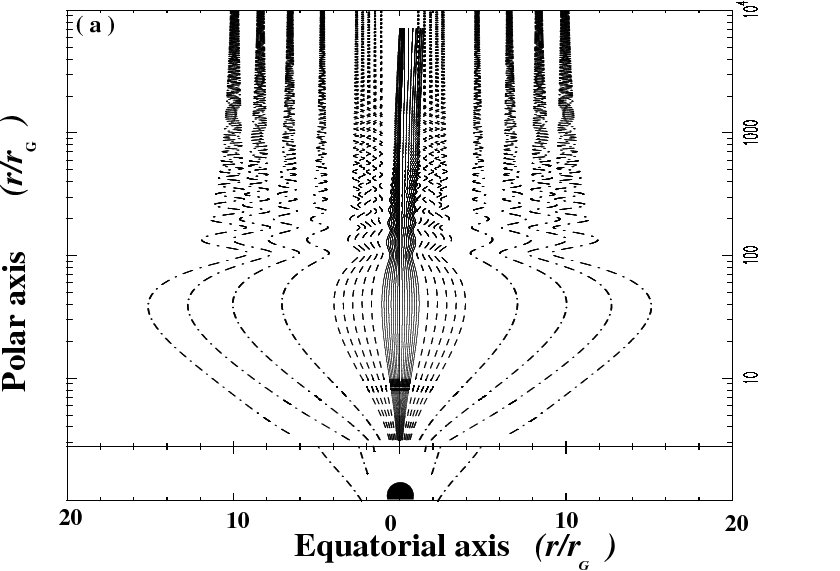}}}}

{\rotatebox{0}{\resizebox{9.0cm}{8.5cm}
{\includegraphics{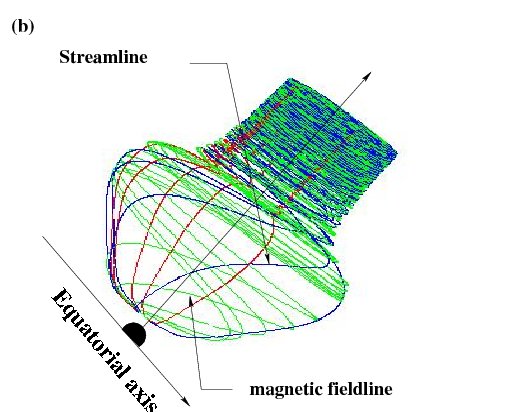}}}}
\caption{Plot of the morphology of the solution corresponding
to a FRI-type  spine jet. In (a) is shown the projection of the
streamlines on the poloidal plane. The solid lines in the center correspond to
lines where the conditions $x_{A}^2 G^2<10^{-2}$ and
$({2 + \delta \alpha})/{\left(1 + \delta \alpha\right)^2} - 2<10^{-2}$
are satisfied. The dashed lines correspond to $x_{A}^2 G^2<10^{-1}$ and
$({2 + \delta \alpha})/{\left(1 + \delta \alpha\right)^2} - 2<10^{-1}$. 
The dashed-dotted lines correspond to
$x_{A}^2 G^2>10^{-1}$ and $({2 + \delta \alpha})/{\left(1 + \delta
\alpha\right)^2} - 2>10^{-1}$ (see Meliani et al. 2006a for details).
\label{Jet_1_application_FRI_Morphologie}}
\end{figure}
\subsection{Temperature}

In polytropic relativistic winds \citep{Melianietal04}, the
temperature is usually defined by the ideal gas equation of state
$P/n = k_{B} T$. Therefore, the knowledge of  specific pressure
$P/n$, the specific thermal energy $e_{\rm th}$ and the density
describe completely the thermodynamics of the fluid (temperature
$T$, enthalpy $w$ and pressure $P$). However, as discussed in
Meliani et al. (2004), flows cannot be adiabatic. The
polytropic approximation is just a convenient way to mimic
heating going on in the flow. Therefore pressure, temperature
and enthalpy are not the real ones but effective quantities that
hide the extra necessary heating.

The temperature definition in meridional self-similar models is
similarly delicate. In fact, as indicated in
 Sauty \& Tsinganos (1999) and Meliani et al. (2006a)
the total gas pressure is
not necessarily limited to be the kinetic pressure. Moreover, the
generalized specific thermal energy of the model, $e_{\rm th} =
\left(w - m c^2 \right)- \displaystyle \frac{P}{n}$, is not
restricted to the thermal energy. In self-similar model these two
quantities are also effective quantities. They account for different
physical processes of energy and momentum transport and dissipation.
They can include the contribution of magnetohydrodynamic waves and
viscous and/or radiative mechanisms. As a matter of fact, the
complexity and variety of MHD processes that can contribute to the
internal energy of a magnetized fluid, makes the definition of the
real thermal energy impossible.

Therefore the quantities $T_{\rm eff}=P/n$ and $e_{\rm th,
eff}=e_{\rm th}$ are simply the specific effective temperature and
thermal energy imposed by the dynamics of the outflow. They do not
necessarily represent neither the kinetic temperature nor the thermal
energy. However they are simple tools to analyze the energetics of
the flow. In the following, we discuss the thermodynamical properties of the
fluid with this effective temperature and not the kinetic
temperature which we cannot calculate.

\begin{eqnarray}\label{Jet_1_result_Temperature_effective}
T_{\rm eff}  = \frac{P}{k_{B} n} = \frac{1}{2 k_{B}}{\gamma_{\star}}^2 w_{\star}
M^2 \Pi \frac{1 + \kappa\alpha}{1 + \delta \alpha} ,
\end{eqnarray}
\begin{eqnarray}\label{Jet_1_result_Thermal_energie_effective}
e_{\rm th,  eff} & = & w - m c^2 -\frac{P}{ n} \nonumber\\
& = &w - m c^2 - \frac{1}{2 k_{B}}{\gamma_{\star}}^2 \frac{w_{\star}}{c^2}
M^2 \Pi \frac{1 + \kappa\alpha}{1 + \delta \alpha} ,
\end{eqnarray}

Thus along the polar axis, the specific thermal energy is deduced from the
Bernoulli equation as follows,
\begin{eqnarray}\label{Jet_1_result_Thermal_energie_effective}
e_{\rm th,  eff}  = w_{\star}  \frac{h_{0, \star}}{h_{0}}
\frac{\gamma}{\gamma_{\star}} - m c^2 - \frac{1}{2}{\gamma_{\star}}^2 w_{\star} M^2 \Pi ,
\end{eqnarray}
We also define ${\cal Q}$ the heat content added to the fluid, which
is the difference between the total effective internal energy of the
fluid $e_{\rm th, eff}$ and the internal energy of the fluid
obtained if it were adiabatic:
\begin{eqnarray}\label{Jet_1_result_heat_energie_effective}
{\cal Q} = e_{\rm th,  eff}  - \sqrt{ m c^2 + \kappa_{\rm pol}
n^{\frac{5}{3} -1} }\,,
\end{eqnarray}
where $\kappa_{\rm pol}$ is a constant.
{ It is
determined from pressure and density at the flow boundary either in
the asymptotic region or in the launching region (in adiabatic flows, we have
$\kappa_{\rm pol}= P / n^{\frac{5}{3}}$).}

\section{Application I - Model of FRI spine jets}\label{FRI}

\subsection{Parameters}

The first solution we show here is adapted to model the spine jet of
radio-loud galaxies of FRI type. The environment of such jets, i.e.
the host galaxy, is known to be rich and containing dense gas.
Moreover, the properties of FRI jets on  the pc scale are quite
different from that on kpc scales. In fact, in the region
close to the nucleus (on the scale of a pc), FRI jets are
accelerated to highly relativistic speeds.

Beyond this region, the jets interact with the external medium which
is denser. This interaction induces an observed deceleration of the
jet. Thus, we assume that the outflow is likely confined by the
pressure of the ambient medium, at least partially. In our model,
these types of jets correspond to solutions associated with inefficient
central magnetic rotators. { In this solution $\eta = 0.90$ which gives an
inefficient magnetic rotator.}

The solution corresponds to the following parameters,
\begin{eqnarray}
\mu& = &0.1 ,\nonumber\\
\nu& = &0.69 ,\nonumber\\
\lambda& = &0.85 ,\nonumber\\
\delta& = & 1.45 ,\\
\kappa& = & 0.42 ,\nonumber\\
\epsilon& = &-0.8475 .\nonumber
\end{eqnarray}

\subsection{Morphology of the FRI-type jet}

As seen in Fig. \ref{Jet_1_application_FRI_Morphologie}, the jet
solution shows an initial expansion up to a distance of $100$
Schwarzschild radii which then stops and the jet recollimates. The
expansion of the jet is due to the strong initial inertia of the
plasma carried along the external streamlines. The jet becomes
collimated once it interacts with the "external" ambient medium
that compresses it. What we call "external" medium refers to
the gas surrounding the last valid streamline of the solution; this
can be the actual external medium of the host galaxy, but
considering the transverse expansion of the solution, it more
likely corresponds to the over-pressured gas of the surrounding
disk wind.

The jet compression for inefficient magnetic rotators
generates strong oscillations in the asymptotic region of the jet
even in the relativistic case \citep{Melianietal06}. The light
cylinder of this jet solution is at infinity. { These oscillations
results from a transfer of energy between the enthalpy and the Lorentz factor as
$h\gamma w$ remains constant to first order with respect to $\alpha$.}

\begin{figure}[th]
{\rotatebox{0}{\resizebox{9.3cm}{4.5cm}
{\includegraphics{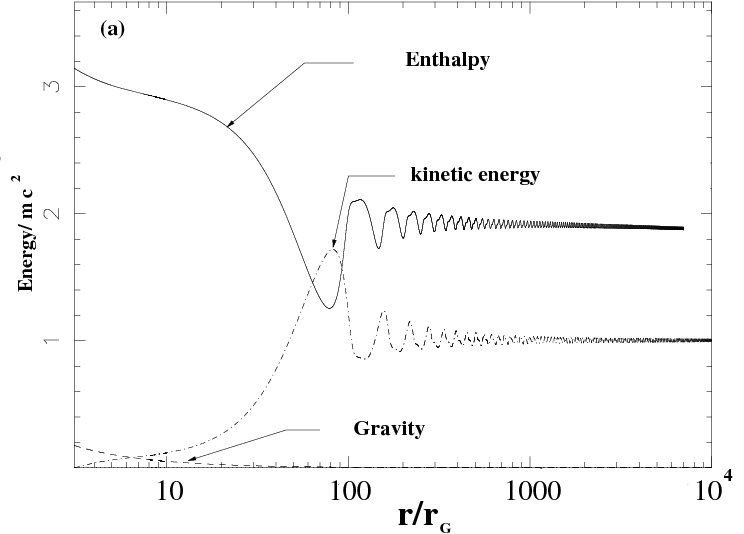}}}}
\\

\rotatebox{0}{\resizebox{9.3cm}{4.5cm}
{\includegraphics{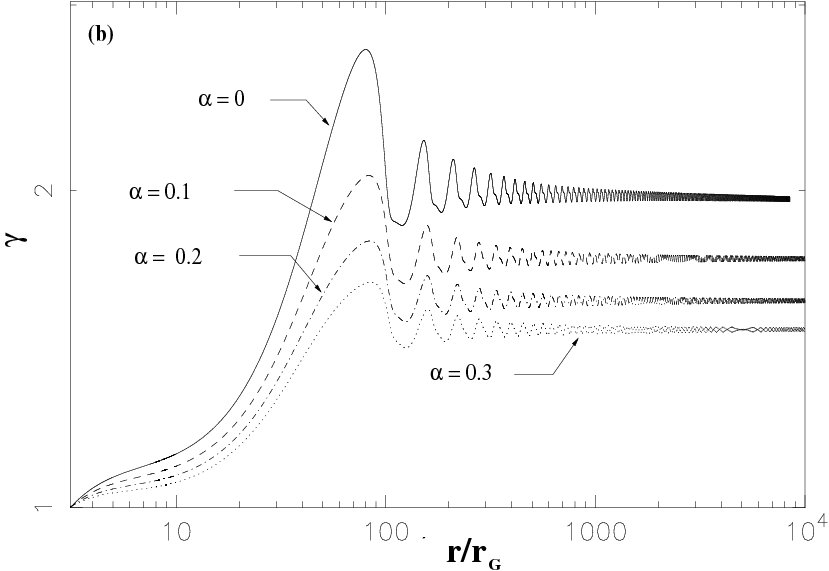}}}
\caption{ In (a) we plot the energetic fluxes normalized to the mass
energy of the first solution for a "FRI-type" spine jet. In (b) we
plot for the same solution, the Lorentz factor $\gamma$ along four
different streamlines. The solid line correspond to the polar axis
and the dotted line to the last streamline connected to the central
corona. Other lines are intermediate ones.}
\label{Jet_1_Application_InEfficace_Energy_acceleration}
\end{figure}

\subsection{FRI-type Jet kinematics}

The jet acceleration occurs mainly in the intermediate region (Fig.
\ref{Jet_1_Application_InEfficace_Energy_acceleration}b) where
gravity becomes weak. Thus, all the enthalpy still remaining in the
outflow is converted to kinetic energy.

{ The maximum Lorentz factor in this solution ($\gamma = 2.8$) is obtained at the largest
expansion of the jet radius, where the  effective temperatures are already ultra relativistic ($\approx
10^{13}$~K).
To explain such a high effective temperature,
part of the pressure must be of non kinetic origin, with contribution from
turbulent ram and magnetic pressure, as we discussed in Sec. 3.1 }
{ Thus this  value of $\gamma$ is close to  the lower observed ones,
but already corresponds to a highly turbulent medium. The model can produce higher
Lorentz factors provided the turbulent pressure level is sufficient. This is
not the main topic of the qualitative discussion we address on the collimation
of the jet itself.
}

In the region of
re-collimation, the increase of the pressure induces a deceleration
of the jet. The Lorentz factor decreases from $\gamma = 2.8$, its
maximum value before the re-collimation, to $\gamma = 2$ in the
asymptotic region. As a matter of fact such a deceleration is indeed a
characteristic of FRI jets as we mentioned. It is remarkable that
this solution shows clearly that the main effect of the
re-collimation by some external pressure is a global deceleration of
the outflow, as observed in FRI jets. The distance of recollimation
in this solution is however smaller than the usual parsec scale.
This may be due to the fact that here we are dealing only with the inner
part of the jet, while observations may correspond to the surrounding disk-wind,
or, it can be due to the fact that our Lorentz factor is too low.
Moreover, we assume here that the external pressure of the host galaxy somehow
is transmitted to the disk-wind, which in turn confines the spine jet.
Keeping in mind these necessary precautions, this result by itself seems interesting,
if we take into account the simplicity of our model.

\subsection{FRI-type Jet energetics}

\begin{figure}[h]
{\rotatebox{0}{\resizebox{9.3cm}{4.5cm}{\includegraphics{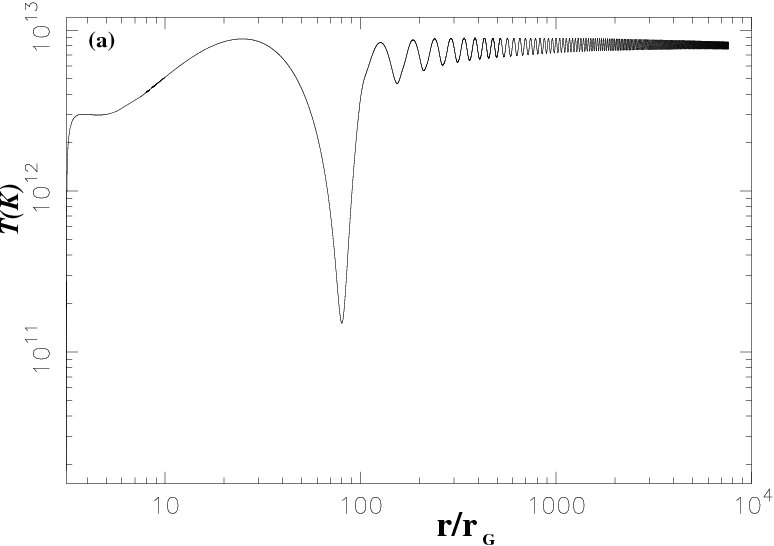}}}}
\\

\rotatebox{0}{\resizebox{9.3cm}{4.5cm}{\includegraphics{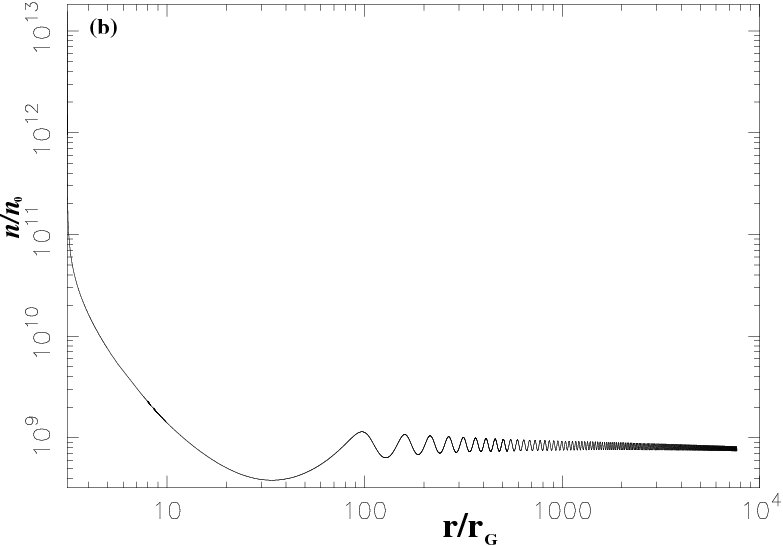}}}
\caption{ In (a) we plot the temperature profile and in (b) the
density profile for the FRI-type jet solution. Density is normalized
to $n_{0}$ such that the mass loss rate is  $\dot{M} = n_{0} 10^{-6} \dot{M}_{\rm Edd}$.}
\label{Jet_1_Application_InEfficace_T_D}
\end{figure}

The temperature profile is characterized by four different regimes.
The three first are common with the "FRII" jet solution and we shall
discuss it later on. The fourth one corresponds to the asymptotic
recollimated region of the "FRI" jet. There, the effective
temperature reaches the high value of $T_{\rm eff} \sim
10^{13}{\rm K}$, because of the strong compression of the outflow by
the external medium (Fig. \ref{Jet_1_Application_InEfficace_T_D}b).
This effective temperature is high compared to the observed
temperature in AGN jets that are usually of the order of $T\sim
10^{8}{\rm K}$. This large difference can be explained however by
some increase of the contribution of non thermal mechanisms to the
effective specific thermal energy $e_{\rm th, eff}$. As we mention before,
we cannot directly compare this effective temperature with the
kinetic temperature in the frame of this model.

\section{Application II - Model for FRII spine jets}\label{FRII}

\subsection{Parameters}

Conversely to the case of FRI, the environment of FRII
radio-loud galaxies is relatively poor. Thus, jets from FRII
galaxies should interact only slightly with the ambient medium. In
these outflows, the velocity increases continuously until the
asymptotic region. In fact, unlike FRI outflows, the velocity in
FRII jets is relativistic both on the parsec and kilo-parsec
scales. Besides, FRII jets are so well collimated on large scales
that they are very likely to have an asymptotic cylindrical shape.
In our model these types of jets correspond to a solution
associated with an efficient central magnetic rotator. Therefore, the value
of the free parameter $\eta$ is probably smaller, i.e. the rotation
is closer to Keplerian velocity than in the case of the solution for FRI.
This increases the available Poynting flux at the base of the jet. Therefore, the jet will be
collimated by the toroidal magnetic pinching without any oscillations.

The parameters for this specific solution are:
\begin{eqnarray}
\mu& = &0.1 ,\nonumber\\
\nu& = &0.65 ,\nonumber\\
\lambda& =  & 1.051 ,\nonumber\\
\delta& = & 1.35 ,\\
\kappa& =  &0.3 ,\nonumber\\
\epsilon& = &0.334 .\nonumber
\end{eqnarray}
{ This is an Efficient Magnetic Rotator. It corresponds to a slightly lower value of $\eta = 0.86$. This
shows that the efficiency is very sensitive to the variations of the rotation frequency. }

\begin{figure}[ht]
{\rotatebox{0}{\resizebox{9.5cm}{10.0cm}
{\includegraphics{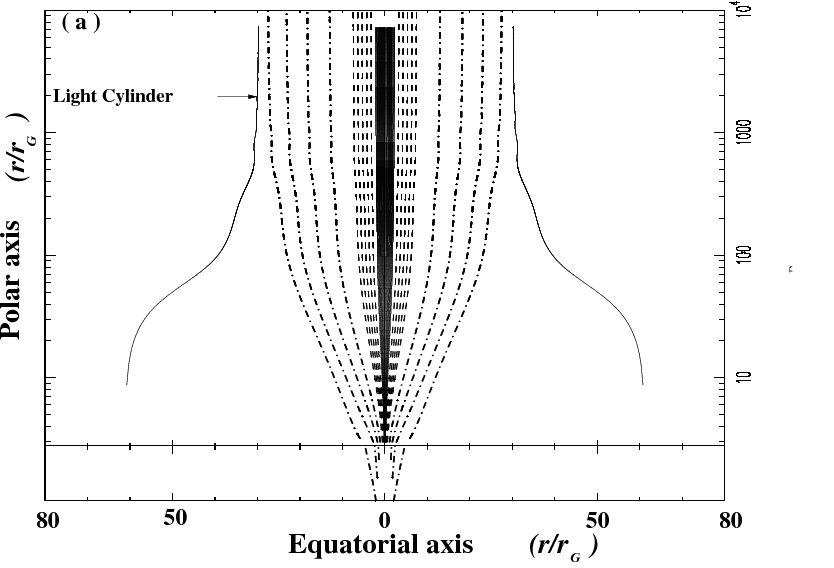}}}}
\vspace{0.1cm}

{\rotatebox{0}{\resizebox{9.0cm}{8.5cm}
{\includegraphics{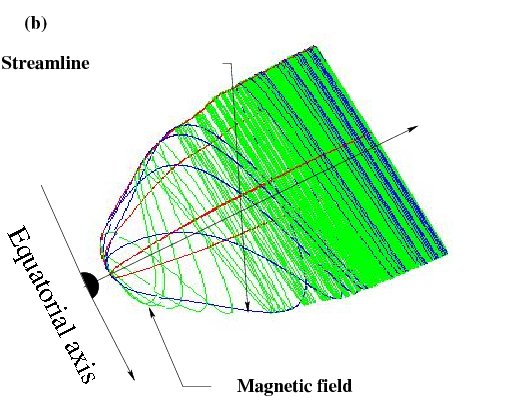}}}}
\caption{The same as in Fig. 3 for the second solution corresponding
to a FRII-type  spine jet.
}
\label{Jet_1_Application_Efficace_Morph}
\end{figure}

This  FRII-type jet solution is characterized by a continuous
expansion up to a distance of about $100$ Alfv\'en radii again, but the
outflow after this distance remains cylindrical, slightly expanding further out.
This expansion is related to the high magnetic pressure at the base,
together with the strong gravity, in addition to some non
negligible contribution of the force of charge separation. However,
the collimation of the jet in the asymptotic region is exclusively
of magnetic origin. It is induced by the toroidal magnetic pinching
force and the transverse magnetic pressure. These two forces balance
the centrifugal and charge separation forces. Oscillations in the
jet are very weak, due  to the relatively
small contribution of the thermal confinement, as expected.

The opening angle of the last open streamline of the solution at a
distance of one parsec { is only $0.1^{\circ}$, } which is rather small
compared to our initial guess. This definitely rules out the
possibility
to describe the whole jet
uniquely with this model. Instead,
we prefer to see it as the spine or inner part of the jet that carries
away the angular momentum of the central black hole. The situation is
similar to what happens in Young Stellar Objects
where the stellar
jet is responsible for the spinning down of the protostar while the outer
disk wind is responsible for the observed mass loss.

\begin{figure}[h]
{\rotatebox{0}{\resizebox{9.3cm}{4.5cm}{\includegraphics{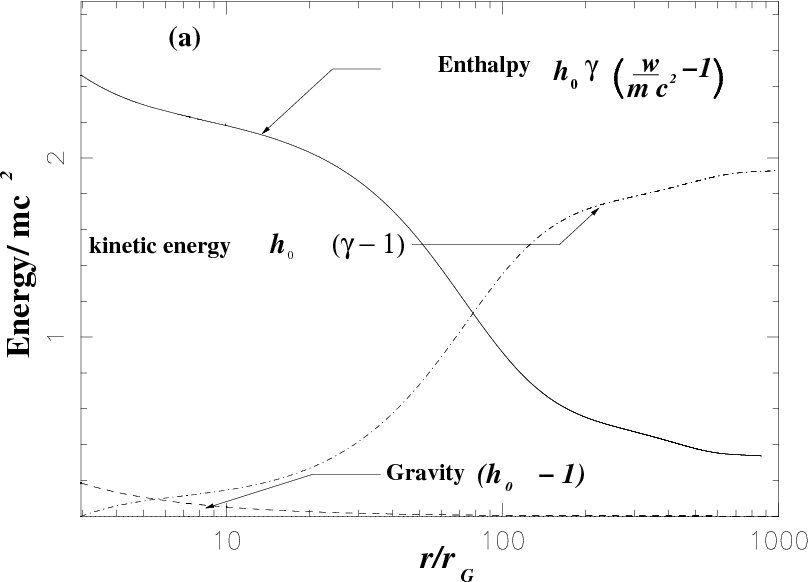}}}}
\\

\rotatebox{0}{\resizebox{9.3cm}{4.5cm}{\includegraphics{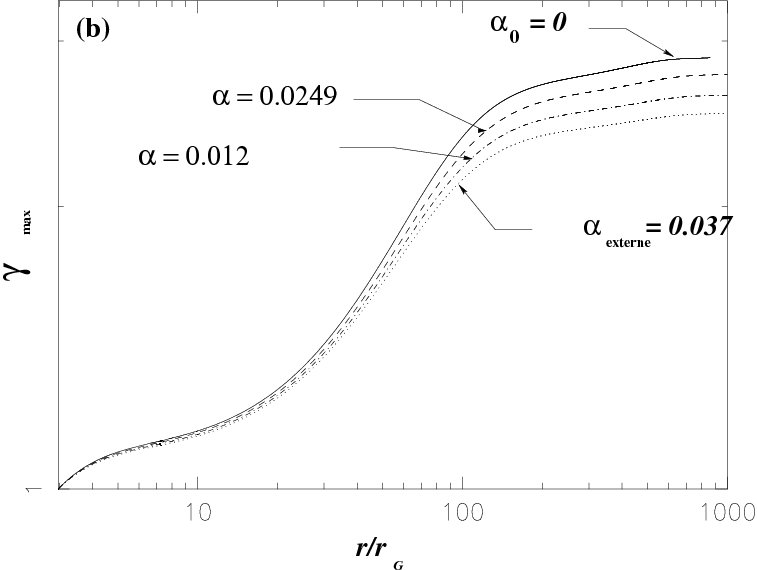}}}
\caption{ In (a) we plot the energetic fluxes normalized to the mass
energy of the second solution for a "FRII-type" spine jet. In (b) we
plot for the same solution, the Lorentz factor $\gamma$ along four
different streamlines. The solid line correspond to the polar axis
and the dotted line to the last streamline connected to the central
corona. Other lines are intermediate ones.}
\label{Jet_1_Application_Efficace}
\end{figure}

\subsection{FRII-type jet kinematics}

The acceleration of this solution is continuous. First, there is a
small but effective thermal acceleration in the lower region of the
corona. In this region, the high thermal energy both accelerates the
fluid up to $0.4$c at a distance of $6 r_{G}$ and enables it to escape
from the deep gravitational potential. A second stronger thermal
acceleration of the jet occurs beyond the Alfv\'en surface up to the
collimation region. In this region, the pressure drops rapidly and
asymptotically goes to negligible values. Therefore, the thermal energy is
transformed into kinetic energy more effectively. In this intermediate
regime the velocity in the flow increases from $0.4$c to $0.92$c on
a scale of the order of $200 r_{G}$.

\subsection{The light cylinder of the FRII-type solution}

{ The light cylinder in the asymptotic region of the jet (Fig.
\ref{Jet_1_Application_Efficace_Morph}) is roughly vertical and asymptotically
parallel to the poloidal streamlines, that  remain inside the
light cylinder}. However, in this solution
conversely to the previous one, the light cylinder is not at infinity
but at a distance of about $20$~$r_{G}$ from the polar axis. This reduces
the domain of validity of the solution around its axis (cf. Meliani
et al. 2006a) where the effects of the light cylinder can be neglected.
{ In fact self similar disk wind models can produce solutions crossing
the light cylinder
 (Vlahakis \& K\"onigl, 2003a,b). Such solutions
undergo a strong magnetic acceleration ideal to obtain high Lorentz factor
in GRBs for instance. This is of course not necessarily the case for the
spine jet which can be accelerated by other means than the
 magnetocentrifugal process.
}

\begin{figure}[h]
{\rotatebox{0}{\resizebox{9.3cm}{4.5cm}
{\includegraphics{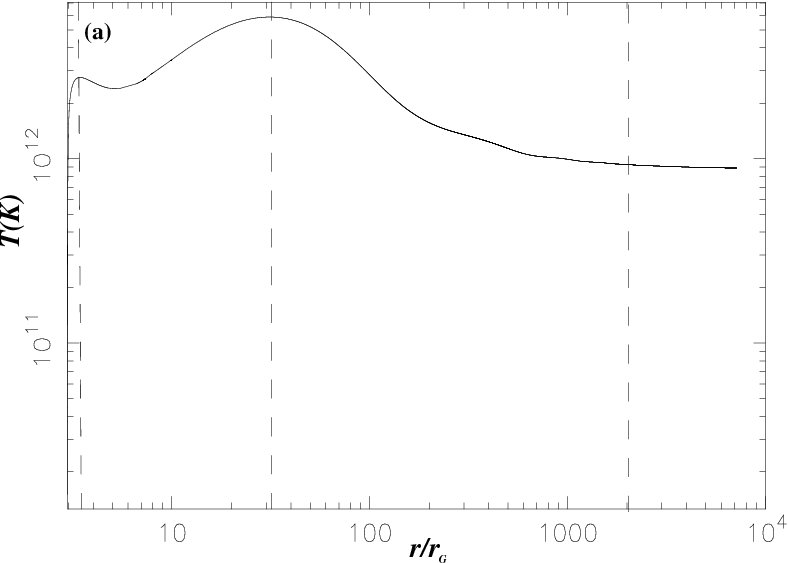}}}}
\\

\rotatebox{0}{\resizebox{9.3cm}{4.5cm}
{\includegraphics{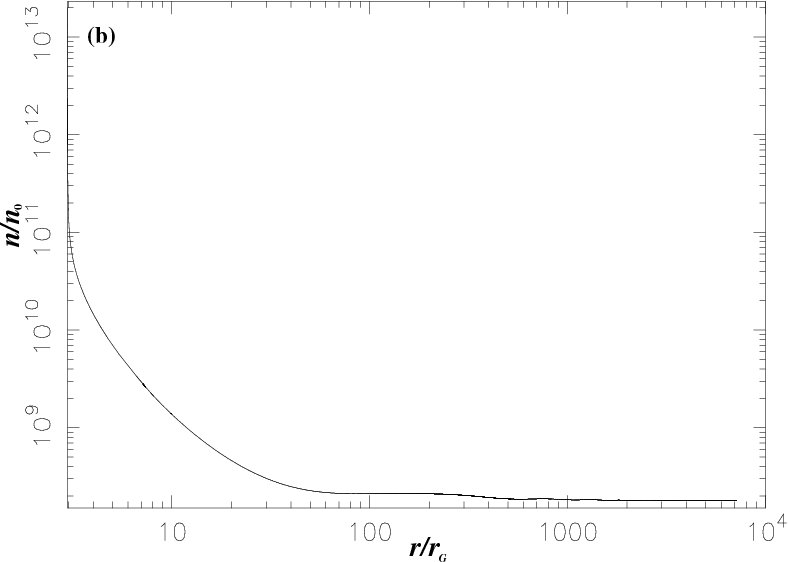}}}
 \caption{ In (a) we
plot the temperature profile and in (b) the density profile for the
FRII-type jet solution. On the left the vertical lines delimitate
the various domains of the temperature profile. On the right,
density is normalized to $n_{0}$ such that the
 mass loss rate is $\dot{M} = n_{0} 10^{-6} \dot{M}_{\rm Edd}$.
}
\label{Jet_1_Application_Efficace_T_D}
\end{figure}

\subsection{FRII-type jet energetics}

The effective temperature profile of this second solution goes through four
different regimes.

First, in the lower corona, the effective temperature increases
extremely rapidly (Fig. \ref{Jet_1_Application_Efficace_T_D}a) from
about $10^{12}{\rm K}$ at the base up to about $3\times10^{12}{\rm K}$. This
increase is
due to some strong initial heating in the expanding corona. The large
expansion induces a strong decrease of the density, but because of the
heating, the pressure decreases less rapidly.

Second, in the intermediate region, the effective temperature still
increases  up to its maximum value of about $3\times10^{12}{\rm K}$, after a
relatively small decrease, because of the global expansion and drop
of the density.

Third, we have a transition region after the maximum and the
asymptotic part. The effective temperature decreases again to attain values around
$10^{12}{\rm K}$. This decrease is induced by the magnetic
compression of the jet that brakes the density decreasing (Fig.
\ref{Jet_1_Application_Efficace_T_D}b).

The effective temperature obtained in this solution, is of course
high compared to observed temperatures in AGN jets which is
typically of the order of $10^{8} {\rm K}$ in the asymptotic
regions. { As in the case of FR I this
indicates that the contribution of non thermal
energies to the acceleration and the heat of the jet play a relevant role.}

\section{Conclusions}\label{conclusions}

We have applied exact GRMHD solutions from
\cite{Melianietal06} to  the  canonical classification of AGN jets
\citep{UrryPadovani95} according to their morphology. Our model is
constructed in the frame of general relativity using the metric of a
central Schwarzschild black hole. It validates the classification of
AGN proposed in Sauty et al. (2001) where the classical MHD
solutions were used  with some additional
features due to relativistic effects. In this study it was proposed that the
inner regions of jets (spines)  are collimated by an
external denser medium in FRI  and  by the force of magnetic
pinching in FRII.

We first proposed a method to estimate the free parameters of
the model from the known properties of AGN jets. In particular, the
departure from Keplerian rotation at the footpoints of the fieldlines is measured by
an extra parameter $\eta$. It turns out that we get very different
classes of solutions by slightly changing this parameter.

First for $\eta = 0.90$
we obtained a
recollimating solution.  On a small spatial scale the outflow expands and
the Lorentz factor reaches a maximum of $\gamma = 2.8$ in the
intermediate region.
This Lorentz factor is obviously smaller by a factor of 2 to 3 from
observed values. We could rescale the whole solution to attain higher
Lorentz factors but at the prize of high temperature in the jet.
Similarly to the Solar wind this would require extra sources for the heating
to account for it. However we were mostly interested in describing the general features
of the solution and try to explain the dichotomy between FRI and FRII jets rather than
modeling specific observed jets.

Then the solution obtained for a typical FRI jet recollimates, undergoes
several oscillations and decelerates down to a Lorentz factor of
$\gamma =2$ in the asymptotic region. This decrease is related to
thermal compression of the jet in the asymptotic region by the outer
medium. We insist on the fact that this recollimation occurs on a
scale smaller than one parsec and that the external pressure is
certainly the pressure of the surrounding disk wind rather than the
external gas from the host galaxy. However if we assume that the
extra pressure of the host galaxy can enhance the pressure in the
disk wind, it is striking to note that our simple toy model for the
spine jet shows a typical feature of FRI jets. Indeed FRI jets show
a deceleration on the kiloparsec scale down to non relativistic
speeds sometimes though they are usually highly relativistic on
smaller scales. Besides that they are also known to have a rich
ambient medium and that the external gas pressure is important as
seen in the X-ray \citep{Capettietal02}. Finally the fact that the FRI
jet radius is larger on the kiloparsec scale can be due precisely to
this recollimation effect which enhances the density such that
probably a larger part of the radio jet is emitting as suggested in
the double component jet of \cite{Soletal89}.

In the case of the FRII jets, the optimal value found is
$\eta =0.86$,
which also gives a maximum Lorentz factor of $\gamma \simeq
3$. The fact that $\eta$ changes very little when we pass from the
"FRI"-type solution to the "FRII"-type, shows that the outflow
properties are highly sensitive to the rotation velocity of the
corona. The solution obtained is however very different as the gas
expands monotonically and remains highly relativistic up to large
distances, as observed in FRII jets. Moreover it is self
collimated by its own magnetic field throughout the length of the jet, something again that is
characteristic of FRII jets wherein there is evidence that the host
galaxy gas is rather underdense.

On the other hand, the fact that in these two cases, $\eta$ is slightly
smaller than  $1$ shows that the central launching region of the
spine jet has to be slightly sub-Keplerian.

In the present paper we did not discuss the case of Seyfert
galaxies which are known to have outflows, though their winds are
not very well collimated and are not relativistic with velocities of
the order of $30,000$ km/s, as observed by the HST. Again, such flows can
be understood in the frame of this simple self-similar model as non
collimated solutions which are radial and exist only if the velocity
does not reach relativistic values \citep{Melianietal06}. Such
solutions are obtained if the magnetic rotator efficiency is very
low, i.e.  for $\epsilon$ very negative, something similar to the
solar wind.

 Altogether then, we may conclude that using a simple toy model for
spine jets, the usual classification of radio sources 
can be understood on one hand by projection effects and Doppler
boosting and on the other by considering the efficiency of the central
magnetic rotator.
 Of course this does not exclude any of the other explanation such as the
 role of the external confinement or shear instabilities. Indeed our FRI-type
 solution is at least partially confined by the
disk wind which may be a signature of external pressure confinement as well.
This idea needs to be further explored with more
sophisticated models or simulations combining the central coronal
jet with an external disk wind, something worth to pursue in another study.

Next step under consideration is the extension of the model to the Kerr metric.
This is a unique chance to construct the first analytical models for a jet around
a rotating black hole. One issue of this extension is to test
the Blandford-Znajek mechanism  (Blandford \& Znajek 1977).

\begin{acknowledgements}
Z. Meliani acknowledges financial support from the FWO, grant G.027708.
 The authors thank Nektarios Vlahakis for helpful discussions and suggestions.

\end{acknowledgements}
\bibliographystyle{aa}

\end{document}